\documentstyle[12pt]{article}

\begin{document}
\title{The masses of vector supermultiplet and of the Higgs supertriplet 
in supersymmetric $SU(5)$ model}
\author{N.V.Krasnikov \thanks{E-mail address: KRASNIKO@MS2.INR.AC.RU}
\\Institute for Nuclear Research\\
60-th October Anniversary Prospect 7a,\\ Moscow 117312, Russia}
\date{August,1996}
\maketitle
\begin{abstract}
The masses of vector supermultiplet and of the Higgs supertriplet in standard 
supersymmetric $SU(5)$ model are calculated. Taking into account 
uncertainties related with the initial coupling constants and threshold 
corrections we find that in standard supersymmetric $SU(5)$ model the scale 
of the supersymmetry breaking could be up to 50 Tev. We find that in the 
extensions of the standard SU(5) supersymmetric model it is possible to 
increase the supersymmetry breaking scale up to $O(10^{12})$ Gev. In standard 
$SU(5)$ supersymmetric model it is possible to increase the GUT scale up to 
$5 \cdot 10^{17}$ Gev provided that the masses of chiral superoctets and 
supertriplets are $m_{3,8} \sim O(10^{13}) Gev$. We also propose SU(5) 
supersymmetric model with 6 light superdoublets and superoctet 
with a mass $O(10^{9})$ Gev.   
\end{abstract}
\newpage

There has recently been renewed interest \cite{1}-\cite{12} in grand 
unification business related with the recent LEP data which allow to measure 
$\sin^{2}(\theta_w)$ with unprecendented accuracy. Namely, the world 
averages with the LEP data mean that the standard nonsupersymmetric $SU(5)$ 
model \cite{13} is ruled out finally and forever (the fact that the 
standard $SU(5)$ model is in conflict with experiment was well known 
\cite{14,15} before the LEP data) but maybe the most striking and impressive 
lesson from LEP is that the supersymmetric extension of the standard 
$SU(5)$ model \cite{16}-\cite{18} predicts the Weinberg angle 
$\theta_w$ in good agreement with experiment. The remarkable success 
of the supersymmetric $SU(5)$ model is considered by many physicists as 
the first hint in favour of the existence of low energy broken supersymmetry 
in nature. A natural question arises: is it possible to invent 
nonsupersymmetric generalizations of the standard $SU(5)$ model 
nonconfronting the experimental data or to increase the supersymmetry 
breaking scale significantly. In the $SO(10)$ model the introduction of 
the intermediate scale $M_I \sim 10^{11} Gev$ allows to obtain the Weinberg 
angle $\theta_w$ in agreement with experiment \cite{19}. In refs.\cite{20,21} 
it has been proposed to cure the problems of the standard $SU(5)$ model 
by the introduction of the additional split multiplets $5 \oplus \overline{5}$ 
and $10 \oplus \overline{10}$ in the minimal $3(\overline{5} \oplus 10)$ of 
the $SU(5)$ model. In ref.\cite{22} the extension of the standard $SU(5)$ 
model with light scalar coloured octets and electroweak triplets has been 
proposed. 

In this paper we discuss the coupling constant unification in standard 
supersymmetric $SU(5)$ model and its extensions. Namely, we calculate 
the masses of two key parameters of SU(5) supersymmetric model - the mass of 
vector supermultiplet and the mass of the Higgs supertriplet. In 
supersymmetric SU(5) model both vector supermultiplet and the Higgs 
supertriplet are responsible for the proton decay. Taking into account 
uncertainties associated with the initial gauge coupling constants 
and threshold corrections we conclude that in standard supersymmetric 
$SU(5)$ model the scale of the supersymmetry breaking could be up to 
50 Tev. We find that in the extensions of the standard $SU(5)$ 
supersymmetric model it is possible to increase the supersymmetry 
breaking scale up to $10^{12}$ Gev. In standard SU(5) supersymmetric model 
it is possible to increase GUT scale up to $5 \cdot 10^{17}$ Gev provided 
that the masses of chiral superoctets and supertriplets $m_{3,8} \sim 
O(10^{13})$ Gev. We also propose SU(5) supersymmetric model with 6 light 
Higgs superdoublets and superoctet with a mass $O(10^{9})$ Gev. 

The standard supersymmetric $SU(5)$ model \cite{16}-\cite{18} contains three 
light supermatter generations and two light superhiggs doublets. 
A minimal choice of massive supermultiplets at the high scale is 
$(\overline{3},2,\frac{5}{2}) \oplus c.c.$ massive vector supermultiplet 
with the mass $M_v$, massive chiral supermultiplets $(8,1,0), (1,3,0), (1,1,0)$ 
with the masses $m_8, m_3, m_1$ (embeded in a 24 supermultiplet of $SU(5)$) 
and a $(3,1,-\frac{1}{3}) \oplus (-3,1,\frac{1}{3})$ complex Higgs 
supertriplet with a mass $M_3$ embeded in $5 \oplus \overline{5}$ 
supermultiplet of $SU(5)$. 
In low energy spectrum we have squark and slepton multiplets 
$ (\tilde{u},\tilde{d})_{L}, \tilde{u}^c_L, \tilde{d}^c_L, 
(\tilde{\nu},\tilde{e})_L, \tilde{e}^c_L $ plus the corresponding 
squarks and sleptons of the second and third generations. 
Besides in the low energy spectrum we have $SU(3)$ octet of gluino 
with a mass $m_{\tilde{g}}$, triplet of $SU(2)$ gaugino with a mass 
$m_{\tilde{w}}$ and the $U(1)$ gaugino with a mass $m_{\tilde{\gamma}}$. 
For the energies between $M_z$ and $M_{GUT}$ we have effective 
$SU(3) \otimes SU(2) \otimes U(1)$ gauge theory. 
In one loop approximation the corresponding solutions of the renormalization 
group equations are well known \cite{18}. In our paper instead of the 
prediction of $\sin^{2}(\theta_w)$ following refs.\cite{6,23,24} we consider the 
following one loop relations between the effective gauge coupling constants,
the mass of the vector massive supermultiplet $M_v$ and the mass of the 
superhiggs triplet $M_3$:
\begin{equation}
A \equiv  2(\frac{1}{\alpha_{1}(m_{t})} - \frac{1}{\alpha_{3}(m_t)}) + 
3(\frac{1}{\alpha_{1}(m_t)} - \frac{1}{\alpha_{2}(m_t)}) = \Delta_{A} ,
\end{equation}
\begin{equation}
B \equiv  2(\frac{1}{\alpha_{1}(m_t)} - \frac{1}{\alpha{3}(m_t)}) - 
3(\frac{1}{\alpha_{1}(m_t)} - \frac{1}{\alpha_{2}(m_t)}) = \Delta_{B} ,
\end{equation}
where
\begin{equation}
\Delta_{A} = (\frac{1}{2\pi})(\delta_{1A} + \delta_{2A} + \delta_{3A}) ,
\end{equation}
\begin{equation}
\Delta_{B} = (\frac{1}{2\pi})(\delta_{1B} + \delta_{2B} + \delta_{3B}) ,
\end{equation}
\begin{equation}
\delta_{1A} = 44ln(\frac{M_v}{m_t}) - 4ln(\frac{M_v}{m_{\tilde{g}}}) -
4ln(\frac{M_v}{m_{\tilde{w}}}) ,
\end{equation}
\begin{equation}
\delta_{2A} = -6(ln(\frac{M_v}{m_8}) + ln(\frac{M_v}{m_3})) ,
\end{equation}
\begin{equation}
\delta_{3A} = 6ln(m_{(\tilde{u},\tilde{d})_L}) - 3ln(m_{\tilde{u}^c_L}) - 
3ln(m_{\tilde{e}^c_L}) ,
\end{equation}
\begin{equation}
\delta_{1B} = 0.4ln(\frac{M_3}{m_h}) + 0.4ln(\frac{M_3}{m_H}) + 
1.6ln(\frac{M_3}{m_{sh}}) ,
\end{equation}
\begin{equation}
\delta_{2B} = 4ln(\frac{m_{\tilde{g}}}{m_{\tilde{w}}}) + 
6ln(\frac{m_8}{m_3}) ,
\end{equation}
\begin{equation}
5\delta_{3B} = -12ln(m_{(\tilde{u},\tilde{d})_L}) + 9ln(m_{\tilde{u}^c_L}) + 
6ln(m_{\tilde{d}^c_L}) - 
6ln(m_{(\tilde{\nu} ,\tilde{e})_L}) + 3ln(m_{\tilde{e}^c_L}) 
\end{equation}
Here $m_h$, $m_H$ and $m_{sh}$ are the masses of the first light Higgs 
isodoublet, the second Higgs isodoublet and the isodoublet of superhiggses.
The relations (1-10) are very convenient since they allow to determine 
separately two key parameters of the high energy spectrum of $SU(5)$ model, 
the mass of the vector supermultiplet $M_v$ and the mass of the chiral 
supertriplet $M_3$. Both the vector supermultiplet and the chiral supertriplet 
are responsible for the proton decay in supersymmetric $SU(5)$ model \cite{18}.
In standard nonsupersymmetric $SU(5)$ model the proton lifetime 
due to the massive vector exchange is determined by the formula \cite{25}
\begin{equation}
\Gamma(p \rightarrow e^{+} \pi^{o})^{-1} = 4 \cdot 10^{29 \pm 0.7}
(\frac{M_v}{2 \cdot10^{14} Gev})^{4} yr
\end{equation}
In supersymmetric $SU(5)$ model the GUT coupling constant is 
$\alpha_{GUT} \approx \frac{1}{25}$ compared to 
$\alpha_{GUT} \approx \frac{1}{41}$ in standard $SU(5)$ model, so we 
have to multiply the expression (11) by factor $(\frac{25}{41})^2$. 
From the current experimental limit \cite{26}
$\Gamma(p \rightarrow e^{+} \pi^{o})^{-1} \geq 9 \cdot 10^{32} yr $ 
we conclude that $M_v \geq 1.3 \cdot 10^{15} Gev$. The corresponding 
experimental bound on the mass of the superhiggs triplet $M_3$ depends 
on the masses of gaugino and squarks \cite{27,28,5}. In our calculations 
we use the following values for the initial coupling constants 
[26,29 - 32]:
\begin{equation}
\alpha_{3}(M_z)_{\overline{MS}} = 0.118 \pm 0.03 ,
\end{equation}
\begin{equation}
\sin^{2}_{\overline{MS}}(\theta_w)(M_z) = 0.2320 \pm 0.0005 ,
\end{equation}
\begin{equation}
(\alpha_{em,\overline{MS}}(M_z))^{-1} = 127.79 \pm 0.13 
\end{equation}
For the top quark mass $m_t = 175 \pm 6$ Gev \cite{33} after the solution 
of the corresponding renormalization group equations in the region 
$ M_z \leq E \leq m_t$ we find that in the $\overline{MS}$-scheme
\begin{equation}
A_{\overline{MS}} = 183.96 \pm 0.47 ,
\end{equation}
\begin{equation}
B_{\overline{MS}} = 13.02 \pm 0.45
\end{equation}
Here the errors in formulae (15,16) are determined mainly by the error in 
the determination of the strong coupling constant $\alpha_{s}(M_{Z})$. 
Since we study the SU(5) supersymmetric model the more appropriate is to 
use the $\overline{DR}$-scheme. The relation between the coupling 
constants in the $\overline{MS}$- and $\overline{DR}$-schemes has the form 
\cite{34}
\begin{equation}
\frac{1}{\alpha_{i_{\overline{MS}}}} = \frac{1}{\alpha_{i_{\overline{DR}}}} 
+ \frac{C_{2}(G)}{12\pi} , 
\end{equation}
where $C_{2}(G)$ is the quadratic casimir operator for the adjoint 
representation. In the $\overline{DR}$-scheme we find that
\begin{equation}
A_{\overline{DR}} = A_{\overline{MS}} + \frac{1}{\pi} = 184.28 \pm 0.47 ,
\end{equation}
\begin{equation}
B_{\overline{DR}} = B_{\overline{MS}} = 13.02 \pm 0.45
\end{equation}

Using one loop formulae (1-10) in the neglection of the contributions  
due to spaticle mass differences and high scale threshold corrections 
($ \delta_{2A} = \delta_{2B} = \delta_{3A} = \delta_{3B} = 0 $ we find that
\begin{equation}
M_{v} = 1.79(\frac{175 Gev}{M_{SUSY}})^{\frac{2}{9}} \cdot 10^{16 \pm 0.04} Gev , 
\end{equation} 
\begin{equation}
M_{3} = 1.1\frac{M_{h,eff}}{175 Gev} \cdot 10^{17 \pm 0.5} Gev ,
\end{equation}
where $M_{SUSY} \equiv (m_{\tilde{g}}m_{\tilde{w}})^{\frac{1}{2}}$ and 
$M_{h,eff} \equiv (m_{h}m_{H})^{\frac{1}{6}}m_{s,h}^{\frac{2}{3}}$.
  
An account of two loop corrections in neglection of the top quark Yukawa 
coupling constant leads to the appearance of the additional factors
\begin{equation}
\delta_{4A,4B} = 2(\theta_1 - \theta_3) \pm 3(\theta_1 - \theta_3) ,
\end{equation}
in the right hand side of the expressions (3,4). Here
\begin{equation}
\theta_{i} = \frac{1}{4\pi}\sum_{j=1}^{3} \frac{b_{ij}}{b_{j}}
ln[\frac{\alpha_{j}(M_v)}{\alpha_{j}(m_t)}]
\end{equation}
and $b_{i}$, $b_{ij}$ are the one loop, two loop $\beta$ function 
coefficients. An account of two loop corrections (22) leads to the increase 
of $M_{v}$ by  factor 1.2 and the decrease of $M_{3}$ by factor 56. 
An account of two loop corrections due to nonzero top quark Yukawa coupling 
constant as it has been found in ref.\cite{11} leads to the additional 
negative corrections to one loop beta function coefficients
\begin{equation}
b_{i} \rightarrow b_{i} - b_{i;top}\frac{h_t^2}{16{\pi}^2} ,
\end{equation}  
where $b_{i;top} = \frac{26}{5}, 6, 4$ for $i = 1, 2, 3$. We have found that 
an account of two loop Yukawa corrections practically does not change the 
value of $M_{v}$ and leads to the small increase of $M_{3}$ by factor 
1.5(1.2) for $h_{t}(m_{t}) = 1$ ($h_{t}(m_{t}) = 0.8$). In the assumption that 
all gaugino masses coincide at GUT scale we find standard relation
$ m_{\tilde{g}} = 
\frac{\alpha_{3}(M_{SUSY})}{\alpha_{2}(M_{SUSY})}m_{\tilde{w}} = 
(2.2 \div 2.5)m_{\tilde{w}} $ between gluino and wino masses that leads 
to the decrease of $M_{3}$ by factor $3.7 \div 4.6$. We have found that the 
values of $\delta_{3B}$ ($\delta_{3A}$)  for realistic spectrum are between 
0 and 0.4 (0 and 5.5) that leads to the maximal decrease of $M_{3}$ 
($M_{v}$) by factor 1.5 (1.4). Taking into account these corrections 
we find that
\begin{equation}
M_{v} = 2.0\cdot(1 \div 0.67)\cdot (\frac{175 Gev}{M_{SUSY}})^{\frac{2}{9}}
\cdot 10^{16 \pm 0.04} Gev ,
\end{equation}
\begin{equation}
M_{3} = 0.80\cdot(1 \div 0.43) \cdot \frac{M_{h,eff}}{175 Gev} 
10^{15 \pm 0.5} Gev 
\end{equation}  
It should be noted that the estimates (25,26) are obtained in the assumption 
$M_{v} = m_{3} = m_{8}$.
 
From the lower bound $1.3 \cdot 10^{15}$ Gev on the value of the mass of 
the vector bosons responsible for the baryon number nonconservation we find 
an upper bound on the value of the supersymmetry breaking parameter 
$M_{SUSY} \leq 1\cdot 10^{8} Gev$. Let us consider now the equations 
(2,8,10,26). For the lightest Higgs mass $m_{h} = 100$ Gev in the assumption 
that $m_{H} = m_{sh} = M_{SUSY}$ and $M_{3} \leq 3M_{v}$ \cite{5} \footnote{The inequality 
$M_{3}  \leq3M_{v}$ comes from the requirement of the absence of Landau pole 
singularities for effective charges for energies up to Planck mass.}
we find that 
\begin{equation}
M_{SUSY} \leq 50 Tev
\end{equation} 
The proton lifetime due to the exchange of the Higgs supertriplet predicts 
much shorter lifetimes for the mode $ p \rightarrow \overline{\nu} K^{+}$ 
than the standard vector boson exchange which leads to the $p \rightarrow 
e^{+} {\pi}^{0}$ proton decay. From the nonobservation 
of $p \rightarrow \overline{\nu} K^{+}$ decay Arnowitt and Nath \cite{5} 
derived an upper limit on the parameter C which can be rewritten in the form
\begin{equation}
C \leq 335 \cdot \frac{M_{3}}{6\cdot10^{16}Gev}\cdot{Gev}^{-1} ,
\end{equation} 
\begin{equation}
C = \frac{-2{\alpha}_{2}}{{\alpha}_{3}\sin(2\beta)}\frac{m_{\tilde{g}}}
{m^{2}_{\tilde{q}}}\cdot 10^{6} {Gev}^{-1}
\end{equation}
From the equations (26,28,29) we find that 
\begin{equation}
\frac{m_{\tilde{g}}}{m^{2}_{\tilde{q}}M_{h,eff}} \leq 84\cdot10^{-9}{Gev}^{-2}
\end{equation}
From the inequality (30) and from the experimental bound \cite{33} 
$ m_{\tilde{g}} \geq 168 Gev$ on the gluino mass in the 
assumption that $m_{\tilde{q}} = m_{H} = m_{sh}$ we find bound on squark mass
\begin{equation}
m_{\tilde{q}} \geq 1460 Gev
\end{equation} 
It should be noted that up to now we assumed that at GUT scale all gaugino 
masses coincide. If we refuse from this requirement it is possible to increase 
the supertriplet mass $M_{3}$ since $M_{3}$ is proportional to 
$(\frac{m_{\tilde{w}}}{m_{\tilde{g}}})^{\frac{5}{3}}$. For instance, for 
$m_{\tilde{g}} = m_{\tilde{w}}$ we find that $M_{3}$ could be up to 
$5.4\cdot10^{16} Gev$ and as a consequence we have more weak bound 
$m_{\tilde{q}} \geq 920 Gev$ for squark mass. Besides for 
$m_{8} \neq m_{3}$ we have additional factor 
$(\frac{m_{3}}{m_{8}})^{\frac{5}{3}}$ in front of the expression for 
the determination of $M_{3}$ that allows to increase the 
value of $M_{3}$.    

It is instructive to consider the supersymmetric $SU(5)$ model with 
relatively light coloured octet and triplets \cite{23}. For instance, 
consider the superpotential
\begin{equation}
W = \lambda \sigma(x)[Tr(\Phi^{2}(x)) -c^2] ,
\end{equation}
where $\sigma(x)$ is the $SU(5)$ singlet chiral superfield and $\Phi(x)$ is 
chiral 24-plet in the adjoint representation. For the superpotential (32) 
the coloured octet and electroweak triplet superfields remain 
massless after $SU(5)$ gauge symmetry breaking and they acquire the 
masses $O(M_{SUSY})$ after the supersymmetry breaking. So in this scenario 
we have additional relatively light fields. Lower bound on the mass 
of the vector bosons leads to the bound on the supersummetry 
breaking scale $M_{SUSY} \leq O(10^{12})$ Gev. 
In order to satisfy the second equation for the mass of the Higgs triplets 
let us introduce in the model two additional superhiggs 5-plets. If we assume 
that after $SU(5)$ gauge symmetry breaking the corresponding Higgs triplets 
acquire mass $O(M_v)$ , the light Higgs isodoublet has a 
mass $O(M_z)$ , the second Higgs isodoublet and superhiggses have masses 
$O(M_{SUSY})$ then we can satisfy the equation (2) for 
$M_{SUSY} \sim 10^{12} Gev$. 

In standard supersymmetric SU(5) model the 
superpotential containing the selfinteraction of the chiral 24-plet 
has the form 
\begin{equation}
W(\Phi(x)) = \lambda[ Tr({\Phi(x)}^3) + MTr({\Phi(x)}^2) ]
\end{equation}
The vacuum solution 
\begin{equation}
\Phi(x) = \frac{4M}{3}Diag(1, 1, 1, -\frac{3}{2}, -\frac{3}{2})
\end{equation}
leads to the $SU(5) \rightarrow SU(3) \otimes SU(2) \otimes U(1)$ gauge 
symmetry breaking. 
After the gauge symmetry breaking the octets and triplets acquire a mass
$m_{8} = m_{3} = 10\lambda M$. So in general superoctet and supertriplet 
masses  don't coincide with $M_{v}$ and in fact they are free parameters of 
the model. It is possible to have grand unification scale 
$M_v =  5\cdot 10^{17} Gev$ and $M_{SUSY} \leq 1 Tev$ \cite{23} provided 
octets and triplets are lighter than the vector supermultiplet by factor 
15000 that is welcomed from the superstring point of view \cite{35}. 

It is interesting to mention that it is possible to have $M_{v} \sim 
5\cdot 10^{17} Gev$ and $M_{SUSY} \leq 1 Tev$ by the introduction only 
relatively light octet with a mass $m_{8} \sim 5\cdot 10^{8} Gev$. To 
satisfy the relation (2) we have to introduce 4 additional light 
superdoublets with masses $ O(10)$ Tev. The phenomenology of such 
models has been considered in ref.\cite{36}. Such models allow to 
have big Yukawa couplings for all generations. The smallness of the 
quark and lepton masses of the first and second generations is related 
with the smallness of the corresponding vacuum expectation values \cite{37}. 
         
In conclusion let us formulate our main results. In standard SU(5) 
supersymmetric model we have calculated the masses of vector 
supermultiplet and of the Higgs supertriplet. We have found that 
in standard supersymmetric $SU(5)$ model with coloured octet 
and triplet masses $O(M_v)$ and with equal gaugino masses at GUT 
scale the nonobservation of the proton decay leads to the upper bound 
$M_{SUSY} \leq 50$ Tev on the  supersymmetry breaking scale and to the 
lower bound $m_{\tilde{q}} \geq 1460$ Gev on the squark mass.
For the case when octets and triplets have the masses $O(M_{SUSY})$ it is 
possible to increase the supersymmetry breaking scale up to 
$O(10^{12}) Gev$ , however in this case in order to satisfy the equation (2) 
for the superhiggs triplet mass we have to introduce 4 additional relatively 
light superhiggs doublets. We have demonstrated also that in standard SU(5) 
model it is possible to have GUT scale $M_{v} \sim 5\cdot 10^{17}$ Gev and 
supersymmetry breaking scale $M_{SUSY} \leq 1$ Tev provided that octets and 
triplets are lighter than vector supermultiplet by factor $O(15000)$.
We have found that it is possible to construct supersymmetric SU(5) model 
with 6 relatively light Higgs superdoublets and the superoctet with a mass 
$m_{8} \sim 5 \cdot 10^{8}$ Gev.  In the extraction of the bound on the 
value of $M_{SUSY}$ our crusial assumption was the inequality \cite{5} 
$ M_3 \leq 3 M_v$. The obtained bound on the $M_{SUSY}$ depends rather 
strongly on the details of the high energy spectrum (on the 
splitting between octet and triplet masses) and on the splitting between 
gaugino masses. It should be noted that for $M_{SUSY} \geq O(1) Tev$ 
we have the fine tuning problem for the electroweak symmetry breaking scale.        

I am indebted to the collaborators of the INR theoretical department 
for discussions and critical comments.

\end{document}